\documentclass[twoside,slac_one]{revtex4}
\usepackage{graphicx}
\usepackage{fancyhdr}
\usepackage{amsmath} 
\usepackage{bm}
\usepackage{amsxtra}
\usepackage{amssymb}
\usepackage{amsthm}
\usepackage{latexsym}
\usepackage{lscape}

\def \met {\mbox{${\not}{E_t}$} }
\begin{document}

\title{Evidence for  $WZ/WW \rightarrow \ell \nu $ + Heavy Flavors Vector Boson Production in 7.5 fb$^{-1}$ of CDF Data}

%

\author{Federico Sforza}
\affiliation{INFN Pisa and Pisa Univeristy, Pisa, IT}

\begin{abstract}

This contribution describes a search for the associated production of Standard Model vector bosons $WZ$ where the $W$
boson decays leptonically ($W\rightarrow \ell \nu$) and the $Z$ boson decays to a heavy flavor quark
pair ($Z \rightarrow b\bar{b},c\bar{c}$). At least one identified (``tagged'') heavy flavor jet in the final state is required.
Given the small di-jet invariant mass separation between the $W$ and the $Z$
resonances, the production of $WW$ events where one $W$
decays leptonically and the second $W$ decays into an heavy flavor jet (e.i. $W^+ \rightarrow c\bar{s}$)
contributes to our the signal.

This search uses data collected with the CDF II detector at the Tevatron Collider at Fermilab, and corresponding to an integrated 
luminosity of approximately 7.5~fb$^{-1}$.
Events consistent with the signature of a charged lepton (electron or muon), large missing transverse energy and exactly two jets, of 
which at least one is required to contain a secondary vertex displaced from the jet
origin, are selected. A multivariate discriminant based on the Support Vector Machine algorithm is used to reduce greatly
the multi-jet background contamination.

We observe a signal of significance of $3.03\sigma$ over the background only hypothesis. A cross section of 1.085 $^{+0.26}_{-0.40}$ 
times the expected Standard Model value for the combined $WZ/WW$ production and decay into heavy flavors is measured,
consistent with the Standard Model prediction.
\end{abstract}

\maketitle

\thispagestyle{fancy}


This contribution describes the search for $p \bar{p} \rightarrow WZ \rightarrow \ell \nu  b \bar{b}$ (or $\rightarrow \ell \nu  c \bar{c}$). 
The signature for this process is a 
$W$-boson, decaying to a high-$P_T$ charged lepton and neutrino, plus a
$Z$-boson decaying to two jets containing heavy flavor quarks. This signature is very similar
to the one used in the search for a low mass Higgs boson ($M_H < 140$~GeV/$c^2$), where the
$H \rightarrow b\bar{b}$ branching fraction is large and the particle is produced in association with a $W$ boson.
Therefore the identification of the $WZ$ signal in the channel containing 
heavy flavor jets represents a benchmark in the search for the low mass Higgs. 

We base our signal to background discrimination on the {\em invariant mass} 
distribution of high-$P_T$ jet pair
entering in our selection, therefore, since we use a secondary vertex finding algorithm to identify $b$-quark produced jets
($b-tagging$), we also consider the process
$WW \rightarrow \ell \nu c\bar{s}$ as part of our signal. We identify
about 8\% of the secondary vertices produced by charmed--hadrons coming from $WW$ events. On the other hand, we identify more than 60\% of 
$WZ$ decaying into heavy flavors.

The main backgrounds for the signal processes include: $W+$jets production (where the jets contain either tagged heavy
flavor or mis--tagged light  flavor), top quark production and multi-jet production, where one jet is misidentified
as a lepton. As we increase the acceptance of our signal using several triggers and lepton identification algorithms,
we use a multi-jet rejection algorithm based on a Support Vector Machine discriminant exploiting the kinematic of the event. 

The CDF detector is described in detail in~\cite{CDF}.

\section{\label{sec:Data}Data Sample \& Event Selection}

 This analysis
is based on an integrated luminosity of 7.5~fb$^{-1}$ collected
with the CDFII detector between March 2002 and March 2011. 
 We select events consistent with the signature of a $W$ boson leptonic decay, large missing transverse energy
and exactly two energetic $b-$quark jets. We accept tight charged lepton candidates, loose charged lepton candidates and isolated tracks;
by construction these lepton categories are orthogonal to each other. 
The data containing tight leptons are collected with an inclusive lepton trigger that requires
an electron (muon) with transverse energy, $E_T$, greater than $18$~GeV (transverse momentum, 
$P_T$, grater than $18$~GeV/c). The data containing loose leptons and isolated tracks are collected 
using triggers based on missing transverse energy (\met) and jet information. 

In the following we refer to these categories of events:
\begin{itemize}
\item CEM: central tight electrons;
\item CMUP and CMX: central tight muons;
\item EMC (extended muon categories): loose muons and isolated track lepton candidates.
\end{itemize}

We select events consistent with a $W$-boson decay plus two energetic b-quark jets. The $W$-boson events are selected
by requiring a single, isolated electron (muon) with $E_T$($P_T$)$>$ 20~GeV(Gev/c) central in the detector (absolute
pseudorapidity in detector coordinates system, $|\eta_{Det}|$, less then $1.1$) and \met $>$ 20 GeV ($>10$ GeV for CMUP and CMX).
Exactly two central ($|\eta_{Det}| < 2.0$) jets with E$_T^{corr}$ $>$ 20 GeV (energy corrected for detector effects) 
are required. In order to improve the separation of signal and background
events, we require that at least one of the two jets is identified to originate from a heavy  quark by the Secondary 
Vertex tagger SecVtx~\cite{secvtx}. We further suppress non-W muti-jet background using a multivariate techinque (see
Section~\ref{sec:SVM})

\subsection{\label{sec:SVM}Suppression of non-W Multi-jet Background}

A fake $W$-boson-like signature can be generated when one jet fakes a high p$_T$ lepton and \met comes from jet energy
mis-measurement.
We developed a method to suppress this, so-called, multi-jet background using a multivariate technique based on the Support
Vector Machine algorithm (SVM)~\cite{SVM}. We developed a software package, based on the LibSVM~\cite{libsvm} library, able to
perform algorithm training, variable ranking, signal discrimination and robustness test.

Although we trained the discriminant on a central electron sample, we apply it to all our selected data-sets (electron,
tight muons, loose muons and tracks) because the algorithm is based only on kinematic variables. We achieve a large reduction
of the multi-jet contamination in all the lepton categories, maintaining a very high efficiency on the signature
$p\bar{p} \rightarrow  \ell \nu  + j j$.

\subsubsection{\label{sec:SVMresults}Summary and Results of the SVM Training}

The test and training sample used to develop the multi-jet veto was built with the following $W$ selection
requirements:
\begin{itemize}
\item one high energy,  isolated central electron;
\item exactly two jets reconstructed with $|\eta_{Det}| <$ 2.0 and $E_T^{corr} >$ 20 GeV;
\item presence of missing transverse energy as signature of the escaping neutrino.
\end{itemize}
Multi-jet events can pass the same requirements, if one of the jets fakes the electron and the \met is either
mis--measured  by the detector, faked by mis--identified (or undetected) minimum ionizing particles or produced by neutrinos
associated with decay of heavy quarks.
We built our training-set using 8000 signal events and 4000 background events (to emulate the data composition):

{\em signal}: W + 2 partons Alpgen Monte-Carlo~\cite{alpgen}, where the $W$ is forced to decay into electron and 
neutrino. We have 
$\approx$ 10$^5$ generated events and we keep $\approx$  9 $\times$  10$^4$ events as a control sample (i.e. not used
for training).

{\em background}: due to the  nature of the background (a mixture of physics processes and detector response), there is
no simulated models that can  be trusted to provide the accurate description needed for training. Therefore we use a
data-driven approach to obtain a  suitable sample: we select events with a fake electron by reversing some of the
``electron quality" requirements (at least 2 out of the 5 cuts),  used to identify the shape of the 
electromagnetic
shower in the calorimeter. This selection is named ``anti-electron'' and produces a multi-jet enriched sample which is, 
however, statistically limited to a few thousands of events. Furthermore, we cannot rely on the modeling of the 
variables directly correlated to the reversed electron cuts.

The variable--sorting algorithm produced an optimal SVM using 6 variables as input features:
\begin{itemize}
\item $W$ related variables: Lepton $P_T$, \met, $\Delta\phi(e,\met^{raw})$;
\item Jet related variables:   $E_T^{raw}$ and  $E_T^{corr}$ of second most energetic jet;
\item Global variables: \met significance (a variable that relates \met with jet
corrections).
\end{itemize}
 
The best SVM configuration reduce the fraction of background in data to $f^{Data}_{bkg}$ $<$ 10\%, and it has an
efficiency on signal events of $\epsilon^{MC}_{sgn}$ $\approx$ 95\% (from MC).

\section{\label{sec:Bak}Backgrounds}

Since our final state has the signature of a charged lepton, \met and two jets (a $W$ boson and jets
signature), the following background sources are considered:
 
{\bf Non-W/Multi-jet} : a W-boson-like signature is generated when one jet fakes a high $P_T$ lepton and \met is  generated
through jet energy mis-measurement (mode details can be found in Section~\ref{sec:SVM}).

{\bf W + Mistags}: this background occurs  when one or more light flavor jets produced in association with a W boson  
are mistakenly identified as a heavy flavor jet by the b-tagging algorithm. Mistags are generated because of the  
finite resolution of the tracking detectors, material interactions, or from long-lived light flavor hadrons
($\Lambda$  and $K_s$) which produce real displaced vertices.

{\bf W+ Heavy Flavor}: these processes ($W + b\bar{b}$, $W + c\bar{c}$ and $W + c$) involve the production of
heavy flavor quarks in association with a $W$ boson.

{\bf Other Electroweak Backgrounds}: additional small but non-negligible background contributions come
from single top quark and top quark pair production, Z boson + jets production.
\\

We determine the amount of selected W+jets events for each lepton category
by fitting the \met distribution of the pretag data control sample: for Top and Electroweak components the MC templates 
are normalized to the theoretical expectation while for W+jets and Non-W the normalization is free to float in the likelihood fit used. 
The following samples are used to produce the non-W templates:
\begin{itemize}
\item modified anti-electrons for central electron  fakes ;
\item non-isolated (iso $>$ 0.2) tight muons for the central tight muons fakes;
\item non-isolated (iso $>$ 0.2) loose  muons to mimic the EMC categories.
\end{itemize}
As expected after the efficient multivariate multi-jet rejection cut, the fits return a very small multi-jet 
contamination (ranging from $2.2$\% to $7.5$\% depending on the lepton category), .

The $b-$tagged $W+$Heavy Flavor ($HF$) component is extracted from the total $W+$jets pretag sample:
the total $W+$jets is composed by a large set of Alpgen+Pythia~\cite{pythia} Monte Carlo weighted by their 
LO production cross section, the $HF$ fractions are then extracted and scaled for the NLO 
contribution and $b-$tagging algorithm efficiency.

We estimate the normalization of W + Mistags background by applying the mistag matrix to the pretag data after
subtracting the non-W, top, diboson, Z+jets and W+HF contributions. We model the W + Mistag kinematics and shapes
using W + Light Flavor Monte Carlo events weighting each event for the mistag probability.

The top quark and other electroweak backgrounds are normalized directly to their theoretical cross sections, calculated
at next-to-leading order.

Finally the residual tagged Non-W component is fitted to the data together with a template of all the other backgrounds: 
the two normalizations are free to float and the multi-jet one is extracted.

More details on the background estimate can be found in Ref~\cite{singletop}. 

Tables~\ref{tbl:pretag},~\ref{tbl:1tag} and ~\ref{tbl:2tags} summarize the number of observed and 
expected events in the W+2 jets sample, for all lepton categories, 
before requiring a b--tag, with one b--tag and with two b--tags, respectively. 

\begin{table}[htb]
\begin{center}
\caption{Summary of pretagged observed and expected events in the W+2 jets sample in 7.5 fb$^{-1}$ of data.}\label{tbl:pretag}
\begin{tabular}{|l|c|c|c|c|c|}
          \hline
          Channel          & CEM         &      CMUP            & CMX           & EMC           & {\bf All Channels} \\ \hline
          Pretag Data       & 61596        &    29036           & 18878         & 27946           &{\bf 137456   } \\ \hline
          $t\bar{t}$        & 498 $\pm$ 46 &    271 $\pm$ 25    & 133 $\pm$ 17  & 418 $\pm$ 39    &{\bf 1320 $\pm$ 95} \\
          Single Top $s$    & 123 $\pm$ 11   &  66 $\pm$ 6      & 32 $\pm$ 4    & 87 $\pm$ 8      &{\bf 308 $\pm$ 22} \\
          Single Top $t$    & 191 $\pm$ 21 &    101 $\pm$ 11    & 52 $\pm$ 7     & 130 $\pm$ 14   &{\bf 474 $\pm$ 37} \\
          WW              & 1580 $\pm$ 132 &    804 $\pm$ 67    & 465 $\pm$ 56   & 822 $\pm$ 70   &{\bf 3672 $\pm$ 253} \\
          WZ              & 216 $\pm$ 20 &      118 $\pm$ 11    & 75 $\pm$ 10    & 147 $\pm$ 14   &{\bf 556 $\pm$ 40} \\
          ZZ              & 4.1 $\pm$ 0.3 &     6.1 $\pm$ 0.5   & 3.7 $\pm$ 0.4  & 8.1 $\pm$ 0.7  &{\bf 22 $\pm$ 2 }\\
          Z+jets                  & 1185 $\pm$ 147 &    1690 $\pm$ 209  & 1084 $\pm$ 163 & 1881 $\pm$ 234 &{\bf 5840 $\pm$ 728} \\
          W + $b\bar{b}$    & 1892 $\pm$ 759 &    905 $\pm$ 362   & 540 $\pm$ 216  & 843 $\pm$ 339  &{\bf 4180 $\pm$ 932} \\
          W + $c\bar{c}$    & 4041 $\pm$ 1622 & 1873 $\pm$ 750  & 1195 $\pm$ 479 & 1724 $\pm$ 693 &{\bf 8833 $\pm$ 1975} \\
          W + cj                  & 3174 $\pm$ 1274 &   1543 $\pm$ 618  & 935 $\pm$ 375  & 1117 $\pm$ 449 &{\bf 6770 $\pm$ 1532} \\
          W +Light Flavor     & 44509 $\pm$ 2785 & 20987 $\pm$ 1104 & 13661 $\pm$ 741 & 18645 $\pm$ 1274 &{\bf 97803 $\pm$ 3339}\\
          Non-W           & 4182 $\pm$ 1673 &   672 $\pm$ 269   & 702 $\pm$ 281  & 2122 $\pm$ 849 &{\bf 7678 $\pm$ 1916} \\
          
          \hline
        \end{tabular}
       \end{center}
\end{table}

\begin{table}[htb]
  \begin{center}
    \caption{Summary of  observed and expected events with one secondary vertex tag (SecVtx), in the W+2 jets sample, in 
      7.5 fb$^{-1}$ of data.}\label{tbl:1tag}
    \begin{tabular}{|l|c|c|c|c|c|}
      \hline
      Chennel & CEM    &      CMUP            & CMX           & EMC           & {\bf All Channels}  \\ 
      \hline 
      Pretag Data       & 61596  & 29036 & 18878  & 27946 &{\bf 137456} \\ \hline
      $t\bar{t}$        &201.3 $\pm$ 19.6   &109.8 $\pm$ 10.7   & 55.0 $\pm$ 7.1     & 171.9 $\pm$ 16.9     &{\bf 538 $\pm$ 53}    \\  
      Single Top $s$    &52.9 $\pm$ 4.8     &28.2 $\pm$ 2.6     & 14.0 $\pm$ 1.8     & 38.0 $\pm$ 3.5       &{\bf  133 $\pm$ 12}    \\  
      Single Top $t$    &71.4 $\pm$ 8.4     &37.4 $\pm$ 4.4     & 19.8 $\pm$ 2.9     & 49.5 $\pm$ 5.8       &{\bf  178 $\pm$ 21 }   \\  
      WW                &68.0 $\pm$ 9.4     &33.3 $\pm$ 4.6     & 20.3 $\pm$ 3.3     & 38.4 $\pm$ 5.3       &{\bf  160 $\pm$ 22  }  \\  
      WZ                &21.8 $\pm$ 2.3     & 11.5 $\pm$ 1.25   & 7.4 $\pm$ 1.0      & 14.1 $\pm$ 1.6       &{\bf  54.7 $\pm$ 5.9   } \\   
      ZZ                &0.44 $\pm$ 0.04    & 0.65 $\pm$ 0.06   & 0.42 $\pm$ 0.05    & 0.86 $\pm$ 0.08      &{\bf  2.4 $\pm$ 0.2 }   \\    
      Z+jets            &27.9 $\pm$ 3.5     & 43.0 $\pm$ 5.5    & 27.3 $\pm$ 4.2     & 65.0 $\pm$ 8.4       &{\bf  163 $\pm$ 21.1}    \\  
      W + $b\bar{b}$    &632.9 $\pm$ 254.2  & 309.8$\pm$ 124.1  & 192.3 $\pm$ 77.1   & 308.9 $\pm$ 124.3    &{\bf  1444 $\pm$ 579}    \\
      W + $c\bar{c}$    &331.0 $\pm$ 133.7  & 155.1 $\pm$ 62.5  & 96.2 $\pm$ 38.8    & 164.2 $\pm$ 66.4     &{\bf  747 $\pm$ 301 }   \\
      W + cj            &259.9 $\pm$ 105.0  & 127.8 $\pm$ 51.5  & 75.3 $\pm$ 30.4    & 106.4 $\pm$ 43.0     &{\bf  569 $\pm$ 229 }   \\
      Mistag            &605.2 $\pm$ 71.3   & 283.8 $\pm$ 31.7   & 181.0 $\pm$ 20.6  & 346.2 $\pm$ 39.2     &{\bf 1416 $\pm$ 146} \\   
      Non-W             &173.9 $\pm$ 69.6   & 45.8 $\pm$ 18.3  & 2.8 $\pm$ 1.1       & 100.9 $\pm$ 40.4     &{\bf  323.3 $\pm$ 129} \\   
      \hline                                                                                                                       
      Prediction        &2446.6 $\pm$ 503.7 & 1186.2 $\pm$ 242.2& 691.8 $\pm$ 148.7  & 1404.5 $\pm$ 242.6   &{\bf  5729 $\pm$ 1132} \\
      Observed          &2332               & 1137              & 699                & 1318                 &{\bf 5486}  \\ 
      \hline                                                                                                                          
      WW/WZ             &89.7 $\pm$ 10.2    & 44.8 $\pm$ 5.05   &  27.7 $\pm$ 3.9   &  52.5 $\pm$ 5.9      &{\bf   214.8 $\pm$ 24.4 }\\
      \hline                                                   
    \end{tabular}  
  \end{center}
\end{table}

\begin{table}[htb]
\begin{center}
\caption{Summary of  observed and expected events with two secondary vertex tags (SecVtx), in the W+2 jets sample, in 
7.5 fb$^{-1}$ of data.}\label{tbl:2tags}
       \begin{tabular}{|l|c|c|c|c|c|}
       \hline
       Chennel & CEM     &      CMUP            & CMX           & EMC           & {\bf All Channels}  \\ 
       \hline 
       Pretag Data       &   61596   & 29036  & 18878  & 27946 &{\bf 137456}\\\hline
       $t\bar{t}$        & 42.2 $\pm$ 6.1  &22.2 $\pm$ 3.2   & 11.1 $\pm$ 1.9  & 34.4 $\pm$ 5.0  &{\bf 109.7 $\pm$ 15.8}  \\
       Single Top $s$    & 14.1 $\pm$ 2.0  &7.6 $\pm$ 1.1    & 3.7 $\pm$ 0.6   & 10.2 $\pm$ 1.4  &{\bf 35.6 $\pm$ 5.0  }  \\ 
       Single Top $t$    & 4.2 $\pm$ 0.7   &2.3 $\pm$ 0.4    & 1.2 $\pm$ 0.2   & 3.1 $\pm$ 0.5   &{\bf 10.8 $\pm$ 1.7 }   \\ 
       WW                & 0.6 $\pm$ 0.1   & 0.26 $\pm$ 0.07 & 0.16 $\pm$ 0.04 & 0.33 $\pm$ 0.08 &{\bf 1.3 $\pm$ 0.3 }   \\  
       WZ                & 4.0 $\pm$ 0.6   & 1.9 $\pm$ 0.3   & 1.4 $\pm$ 0.2   & 2.4 $\pm$ 0.4   &{\bf 9.6 $\pm$ 1.4 }   \\  
       ZZ                & 0.06 $\pm$ 0.01 & 0.12 $\pm$ 0.02 & 0.09 $\pm$ 0.01 & 0.16 $\pm$ 0.02 &{\bf 0.43 $\pm$ 0.06 }   \\
       Z+jets            & 0.9 $\pm$ 0.1   &2.0 $\pm$ 0.3    & 1.2 $\pm$ 0.2   & 3.1 $\pm$ 0.4   &{\bf 7.2 $\pm$ 1.0   } \\    
       W + $b\bar{b}$    & 81.9 $\pm$ 33.2 &42.2 $\pm$ 17.1  & 23.4 $\pm$ 9.5  & 44.9 $\pm$ 18.2 &{\bf 192 $\pm$ 78}  \\
       W + $c\bar{c}$    & 4.7 $\pm$ 1.9   &2.3 $\pm$ 1.0    & 1.3 $\pm$ 0.5   & 2.8 $\pm$ 1.1   &{\bf 11.0 $\pm$ 4.5 }   \\ 
       W + cj            & 3.7 $\pm$ 1.5   &1.9 $\pm$ 0.8    & 1.0 $\pm$ 0.4   & 1.8 $\pm$ 0.7   &{\bf 8.3 $\pm$ 3.4  }  \\    
       Mistag            & 3.2 $\pm$ 0.7   &1.6 $\pm$ 0.3    & 0.9 $\pm$ 0.2   & 2.2 $\pm$ 0.4   &{\bf 7.8 $\pm$ 1.6  }  \\  
       Non-W             & 7.9 $\pm$ 3.2   &4.8 $\pm$ 1.9    & 0.1 $\pm$ 0.5   & 0.0 $\pm$ 0.5   &{\bf 12.8 $\pm$ 6.1 }   \\   
       \hline                                                                                                           
       Prediction        & 167.3 $\pm$ 38.0&88.9 $\pm$ 19.6  & 45.4 $\pm$ 10.9 & 105.3 $\pm$ 21.5 &{\bf 406.9 $\pm$ 89.5}\\
       Observed          & 147             &74               & 39              & 106             &{\bf 366}    \\              
       \hline                                                                                                           
       WW/WZ             & 4.6 $\pm$ 0.6   & 2.1 $\pm$ 0.3   & 1.5 $\pm$ 0.2   & 2.7 $\pm$ 0.4   &{\bf   10.9 $\pm$ 1.5} \\
       \hline
       \end{tabular}  
\end{center}
\end{table}

\section{M$_{inv}(jet1jet2)$ distribution}
        
\begin{figure} [ht]
\centering
\includegraphics[width=80mm]{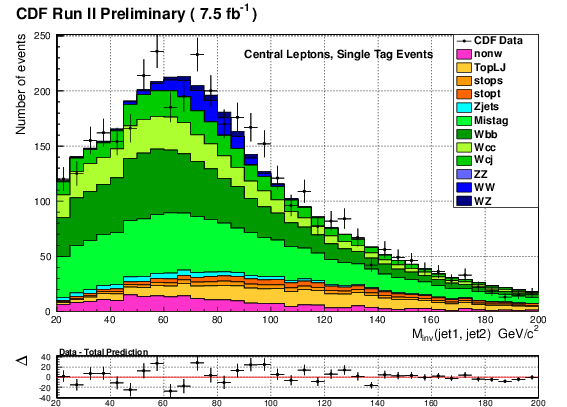}
\caption{\em $M_{Inv}(jet1,jet2)$ distribution for the 1 SecVtx tag candidates, tight leptons (CEM+CMUP+CMX combined). The best fit of 
the systematic nuisance parameters are taken into account.}  
\label{fig:mjj_1tag_cen}
\end{figure}

\begin{figure} [ht]
\centering
\includegraphics[width=80mm]{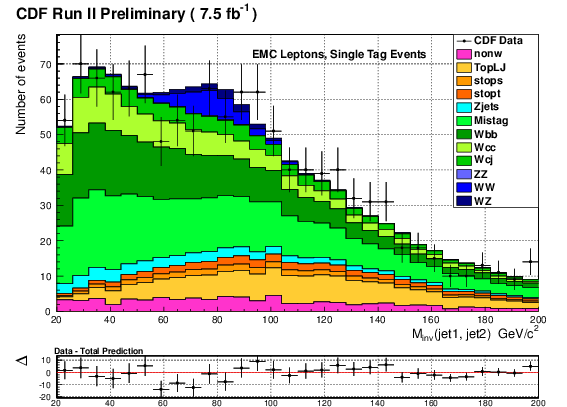} 
\caption{\em $M_{Inv}(jet1,jet2)$ distribution for the 1 SecVtx tag candidates, extended muon categories (EMC). The best fit of
the systematic nuisance parameters are taken into account.}
\label{fig:mjj_1tag_emc}
\end{figure}

\begin{figure} [ht]
\centering
\includegraphics[width=80mm]{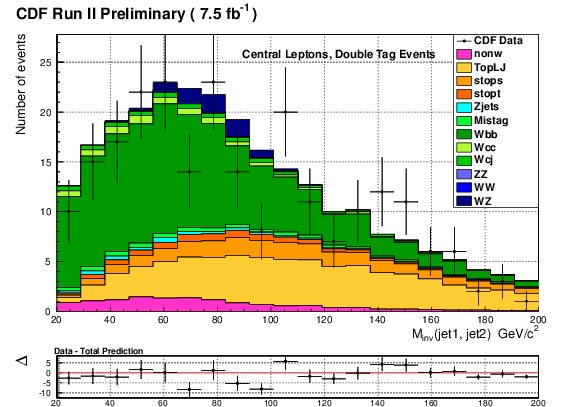}
\caption{\em $M_{Inv}(jet1,jet2)$ distribution for the 2 SecVtx tag candidates, tight leptons (CEM+CMUP+CMX combined). The best fit of
the systematic nuisance parameters are taken into account. } 
\label{fig:mjj_2tag_cen}
\end{figure}

\begin{figure} [ht]
\centering
\includegraphics[width=80mm]{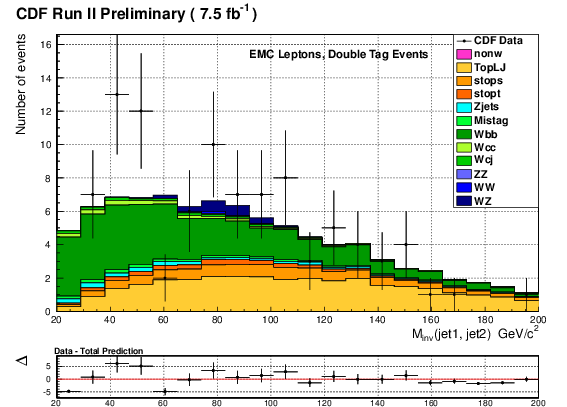}
\caption{\em $M_{Inv}(jet1,jet2)$ distribution for the 2 SecVtx tag candidates, extended muon categories (EMC). The best fit of
the systematic nuisance parameters are taken into account. }
\label{fig:mjj_2tag_emc}
\end{figure}

The signal discrimination is based on the invariant mass of the two jets (M$_{inv}(jet1jet2)$) in the event. Candidates are separated into four 
statistically independent channels: tight leptons (CEM+CMUP+CMX) with 1 SecVtx tag, EMC lepton candidates with 1  SecVtx tag, 
tight leptons with 2 SecVtx tags and EMC with 2 SecVtx tags.
The M$_{inv}(jet1jet2)$
distributions for single tagged events for tight leptons and EMC leptons 
are shown in Figure~\ref{fig:mjj_1tag_cen} and Figure~\ref{fig:mjj_1tag_emc} respectively. The corresponding 
distributions for events
 with two tags are shown in  Figure~\ref{fig:mjj_2tag_cen} and Figure~\ref{fig:mjj_2tag_emc}. These four distributions are used for the 
final signal to background discrimination. The M$_{inv}(jet1jet2)$ plots shown here are the ones returned by the final fit, with a full
treatment of the correlated systematic effects 
(see next paragraph for a complete description).

\section{Statistical Analysis}

Since the process $WZ/WW \rightarrow \ell \nu $ + Heavy Flavors was never observed before,
we start evaluating 95\% C.L. limits on a potential signal. The following systematic uncertainties 
(for background and signal) are taken 
into account as normalization nuisance parameters: JES, Alpgen $Q^2$, b-tag scale factor,
lepton identification and trigger efficiencies, multi-jet background normalization, 
NLO scaling of W+heavy flavor production, ISR/FSR (for signal only) and
mistag uncertainty. In addition, JES and $Q^2$ are taken as shape systematics as well, where the
interpolated shape variation is used as nuissance parameter. All the nuissance parameters are fitted
to improve the sensitivity. 

Expected 95\% C.L. limits, assuming no SM $WW/WZ$ production, are determined using Monte Carlo pseudo experiments based on expected
yields varied within the assigned systematics. The normalization and shape uncertainties are integrated into the limit calculations.
Table~\ref{tab:limits} summarizes the median expected (and observed) 95\% production limits  (in units of SM $WW+WZ$ cross 
section multiplied 
by branching fraction into heavy flavors).

\begin{table}
\begin{center}
\caption{Expected limits for each lepton category, single and double tagged events, in units of the SM cross section for $WW+WZ$ production
multiplied by the branching fraction into heavy flavors.}
\begin{tabular}{|l|c|c|}
\hline
Tag and lepton category              & Expected limit & Observed limit\\ \hline
1 tag Tight Leptons (CEM+CMUP+CMX) & 0.72 $^{+0.42}_{-0.38}$ & 2.07              \\
1 tag EMC                            & 1.21 $^{+0.57}_{-0.46}$ & 1.70              \\
2 tags Tight Leptons (CEM+CMUP+CMX)& 4.02 $^{+1.94}_{-1.50}$ & 3.03               \\
2 tags EMC                           & 6.07 $^{+2.92}_{-2.20}$ & 8.59               \\ \hline
All combined                         & 0.57 $^{+0.33}_{-0.31}$ & 1.46               \\ \hline                                                                                                                                          
\end{tabular}
\end{center}
\label{tab:limits}
\end{table}

Combining single--tagged and double--tagged results for all lepton categories,  we find an expected limit of:
0.575 $^{+0.33}_{-0.31}$ times the SM prediction.

The observed limit, combining single--tagged and double--tagged results for all lepton categories, is 1.46
times the SM prediction.

\subsection{Sensitivity}

\begin{figure} [ht]
\centering
\includegraphics[width=80mm]{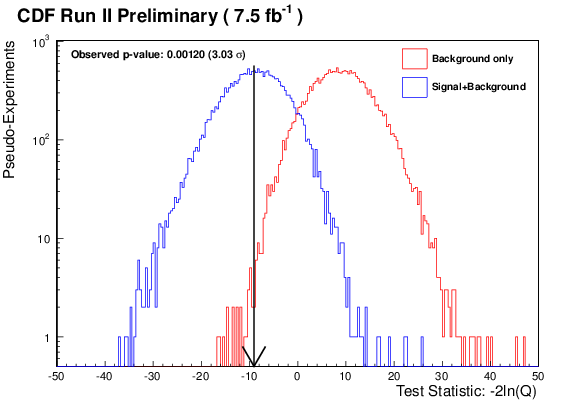}
\caption{\em Distributions  of $ - 2 ln Q$  for the test hypothesis $H_1$, which assumes Standard Model
 backgrounds plus 
Standard-Model $WW+WZ$ production and decay into heavy flavors (blue histogram), and for the null 
hypothesis, $H_0$, 
which assumes no $WW+WZ$  (red
histogram). The observed value of $ - 2 ln Q$  is indicated with a solid, vertical line. 
The plot is shown on a logarithmic scale.
The p-value is the fraction of the integral of the $H_0$ curve to the left of the data.}
\label{fig:gaussians}
\end{figure}

To compute the significance of a potentially observed signal, we perform a hypothesis test, comparing the data to two hypotheses.
The null hypothesis, $H_0$, assumes Standard Model processes except $WW+WZ$ production. The second hypothesis, $H_1$, 
assumes that the $WW+WZ$ production cross  section and the branching ratio into heavy flavors are  the ones
 predicted by the Standard Model. The likelihood ratio is defined as:
\begin{eqnarray}
-2 ln Q = -2 ln \frac{p(data|H_1,\hat{\theta})}{p(data|H_0,\hat{\hat{\theta}})}
\end{eqnarray}
where $\theta$ represents the nuisance parameters describing the uncertain values of the quantities studied for systematic error,
$\hat{\theta}$ the best fit values of $\theta$  under $H_1$ and $\hat{\hat{\theta}}$ are the best fit 
values of the nuisance parameters under $H_0$.
We perform two sets of pseudo-experiments to determine the expected sensitivity to the signal~\cite{mclimit}, one assuming $H_0$ and a second one 
assuming $H_1$. On each pseudoexperiment, the values of the nuisance parameters are chosen randomly based on
the systematic errors. The distributions of the values of $ - 2 ln Q$ are shown in Fig.~\ref{fig:gaussians} for the two hypotheses and the data.

The $p-$value is the probability that $ - 2 ln Q < -2ln Q_0$, assuming the null hypothesis $H_0$.
The $p-$value was found to be 0.00120, corresponding to a 3.03$\sigma$ excess.  

The sensitivity of the analysis is computed as the median
expected $p-$value assuming a signal is truly present. The median $ - 2 ln Q$ is extracted from the $H_1$ 
distribution, and 
the integral of the $H_0$ distribution of $ - 2 ln Q$  to the left of this median value is the median expected $p$-value. 
The value obtained is 0.00126, corresponding to 3.02$\sigma$.

\subsection{$WW+WZ$ cross section measurement}

\begin{figure} [ht]
\centering
\includegraphics[width=80mm]{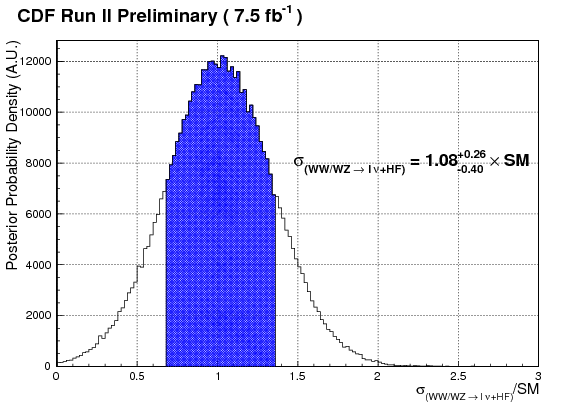}
\caption{\em The Bayesian posterior,
marginalized over nuisance parameters, is shown. The maximum value is the central value of the cross-section. The blue area represents the 
 smallest interval enclosing 68\% of the integral of the posterior.}
\label{fig:posterior}
\end{figure}

In order to measure the $WW+WZ$ production cross section, a Bayesian marginalization technique is applied to the 
$M_{inv}(jet1,jet2)$ distribution in both 1 tag and 2 tags samples. The nuisance parameters are integrated out as 
described in
\cite{mclimit}. The distribution of the posterior is shown in Fig.~\ref{fig:posterior}. The 
maximum of the posterior is taken to be the best fit value for the cross section measurement, and the 1-$\sigma$ confidence interval 
is taken to be the shortest interval containing 68\% of the integral of the posterior
distribution. The resulting cross section measurement, in units of expected SM signal, is:
1.085$^{+0.26}_{-0.40}$. 

\section{Conclusion}
We analyzed $7.5$~fb$^{-1}$ of data looking for the $WW/WZ\rightarrow \ell \nu+ HF$ signal in
the $W+2$ jets exclusive sample.
We found an excess over background in the $M_{inv}(jet1,jet2)$ distribution looking at
the Double tagged + Single tagged samples.
Using a background only-hypothesis we found our result inconsistent 
with data: the significance of the observed signal corresponds to $3.03 \sigma$.
We performed a measurement of the cross section for this process, that
we found to be 1.085 $^{+0.26}_{-0.40}$ times the expected Standard Model prediction.

\begin{acknowledgments}
We thank the Fermilab staff and the technical staffs of the
participating institutions for their vital contributions. This
work was supported by the U.S. Department of Energy and National
Science Foundation; the Italian Istituto Nazionale di Fisica
Nucleare; the Ministry of Education, Culture, Sports, Science and
Technology of Japan; the Natural Sciences and Engineering Research
Council of Canada; the National Science Council of the Republic of
China; the Swiss National Science Foundation; the A.P. Sloan
Foundation; the Bundesministerium fuer Bildung und Forschung,
Germany; the Korean Science and Engineering Foundation and the
Korean Research Foundation; the Particle Physics and Astronomy
Research Council and the Royal Society, UK; the Russian Foundation
for Basic Research; the Comision Interministerial de Ciencia y
Tecnologia, Spain; and in part by the European Community's Human
Potential Programme under contract HPRN-CT-20002, the Slovak R and D agency and the Academy of Finland.
\end{acknowledgments}

\bigskip 

\end{document}